\documentclass[useAMS,usenatbib]{mn2e}
\usepackage{times}
\usepackage{graphics,epsfig}
\usepackage{graphicx}
\usepackage{amssymb}

\title[Accretion in fractal media]{On spherically symmetrical accretion in fractal media}
\author[N. Roy]{Nirupam Roy \thanks{E-mail: nirupam@ncra.tifr.res.in}\\ 
       NCRA-TIFR, Post Bag 3, Ganeshkhind, Pune 411 007, India}

\begin{document}
\date{Accepted yyyy mmm dd. Received yyyy mmm dd; in original form yyyy mmm dd}

\pagerange{\pageref{firstpage}--\pageref{lastpage}} \pubyear{yyyy}

\maketitle

\label{firstpage}

\begin{abstract}

We use fractional integrals to generalize the description of hydrodynamic 
accretion in fractal media. The fractional continuous medium model allows the 
generalization of the equations of balance of mass density and momentum density. 
These make it possible to consider the general case of spherical hydrodynamic 
accretion onto a gravitating mass embedded in a fractal medium. The general 
nature of the solution is similar to the ``Bondi solution'', but the 
accretion rate may vary substantially and the dependence on central mass may 
change significantly depending on dimensionality of the fractal medium. The 
theory shows consistency with the observational data and numerical simulation 
results for the particular case of accretion onto pre-main-sequence stars.

\end{abstract}


%%%%%%%%%%%%%%%%%%%%%%%%%%%%%%%%%%%%%%%%%%%%%%%%%%%%%%%%%%%%%%%%%%%%%%%%%%%%%%%
%%%%%%%%%%%%%%%%%%%%%%%%%%%%%%%%%%%%%%%%%%%%%%%%%%%%%%%%%%%%%%%%%%%%%%%%%%%%%%%
\section{Introduction}
%%%%%%%%%%%%%%%%%%%%%%%%%%%%%%%%%%%%%%%%%%%%%%%%%%%%%%%%%%%%%%%%%%%%%%%%%%%%%%%
%%%%%%%%%%%%%%%%%%%%%%%%%%%%%%%%%%%%%%%%%%%%%%%%%%%%%%%%%%%%%%%%%%%%%%%%%%%%%%%
\label{sec1}
The interstellar medium (ISM) is believed to have a self-similar hierarchical 
structure over several orders of magnitude in scale \citep*{lars,falg92,heith}. 
Direct H~{\sc i} absorption observations and interstellar scintillation 
measurements suggest that the structure extends down to a scale of 10 AU 
\citep*{crov,lang,fais} and possibly even to a scale of sub-AU \citep*{hill}. 
However, the latter is limited by the spatial resolution of the observations. 
Hence the issue is far from being definite even after observational detection 
of lower limit of self-similarity scale in some ISM components \citep*{good}. 
Numerous theories have attempted to explain the origin, evolution and mass 
distribution of these clouds \citep*[began with the hierarchical fragmentation 
picture,][]{hoyle} and it has been established, from both observations 
\citep*{elme} and numerical simulations \citep*{burk,kless,seme}, that the 
interstellar medium has a clumpy hierarchical self-similar structure with a 
fractal dimension $2.5 \lesssim D \lesssim 2.7$ \citep*{nestor} in 3 
dimensional space. The main reason for this is still not properly understood 
but can result from the underlying fractal geometry that may arise due to 
turbulent processes in the medium.

Here we have investigated physical processes, in particular hydrodynamics, in 
such a medium with fractal dimension. We have considered the simplest situation 
of hydrodynamic spherical steady accretion of the medium onto a large gravitating 
mass embedded in this medium. We have assumed the dimensionality of the medium 
to be isotropic and homogeneous. It implies that the mass density (or, 
equivalently, dimensionality) is same for any surface independent of the 
orientation. The famous solution of the continuous medium case is the ``Bondi 
Solution'' for spherical hydrodynamic accretion \citep*{bondi}. Here we have 
extended that analysis for a medium with fractal dimension $D~(=3d)\leq 3$.

To describe physical processes in fractal medium, we have used fractional 
integration and differentiation \citep*[][and references therein]{zas}. We 
replace the fractal medium by some continuous medium and the integrals on the 
network of fractal medium is approximated by fractional integrals \citep*{ren}. 
The interpretation of fractional integration is connected with fractional mass 
dimension \citep*{mand}. Fractional integrals can be considered as integrals 
over fractional dimension space \citep*[up to a numerical factor;][]{tara}. We 
have chosen the numerical factor properly to get the right dimension of any 
physical parameter and to derive the standard expression in the limit $d 
\rightarrow 1$. The isotropic and homogeneous nature of dimensionality is also 
incorporated properly. This model allows us to describe the hydrodynamics in a 
self-consistent way in a fractal medium.

In this paper we first derive, in \S2, the steady state mass accretion rate for 
fractal spherical hydrodynamic accretion using simple dimensional analysis. In 
\S3 we derive the steady state hydrodynamic equations to describe the spherical 
accretion in a fractal medium and the steady state accretion rate in terms of 
boundary conditions at infinity is derived in \S4. We discuss the actual 
astrophysical implication of our analysis and compare our central result with 
observational data and numerical simulation results in \S5. Possible limitations 
of our analysis is discussed in \S6. Finally, we summarize and present our 
conclusions in \S7.

%%%%%%%%%%%%%%%%%%%%%%%%%%%%%%%%%%%%%%%%%%%%%%%%%%%%%%%%%%%%%%%%%%%%%%%%%%%%%%%
%%%%%%%%%%%%%%%%%%%%%%%%%%%%%%%%%%%%%%%%%%%%%%%%%%%%%%%%%%%%%%%%%%%%%%%%%%%%%%%
\section{Dimensional Analysis of Fractal Accretion}
%%%%%%%%%%%%%%%%%%%%%%%%%%%%%%%%%%%%%%%%%%%%%%%%%%%%%%%%%%%%%%%%%%%%%%%%%%%%%%%
%%%%%%%%%%%%%%%%%%%%%%%%%%%%%%%%%%%%%%%%%%%%%%%%%%%%%%%%%%%%%%%%%%%%%%%%%%%%%%%
\label{sec2}
We assume a medium that, on a range of length scale $R$, has a fractal structure 
of dimensionality $D=3d<3$ embedded in 3 dimensional space. Here $D$ refers to 
the mass dimension and it implies that in such a medium of constant density 
$\rho$, the mass enclosed in a sphere of radius $r$ will be 
\begin{equation}
M_D = kr^D \sim \frac{\rho}{l_c^{3(d-1)}}R^{3d}
\end{equation}
where $l_c$ is a characteristic inner length of the medium and can take arbitrary 
value in the limit $d \rightarrow 1$. It is the scale below which the medium 
will be continuous. We can define the modified ``density'' for this fractal 
medium as $\overline{\rho}\equiv\rho/l_c^{3(d-1)}$ so that $M_D\sim\overline
{\rho}R^{3d}$.

For steady state hydrodynamic spherical accretion onto a central mass $M$ from 
its surroundings fractal medium with a mass dimension of $D$, the relevant 
physical parameters will be (1) sound speed at a large distance away from the 
accretor ($a_\infty$), (2) modified ``density'' of the fractal medium at a large 
distance away from the accretor ($\overline{\rho}_\infty=\rho_\infty/l_c^
{3(d-1)}$) and (3) mass of the accretor scaled by the gravitational constant 
($GM$). The dimensions of these three parameters are
\begin{equation}
[a_\infty]=[M]^{ 0}[L]^{ 1}[T]^{ 1}
\end{equation}
\begin{equation}
[\overline{\rho}_\infty]=[M]^{ 1}[L]^{-D}[T]^{ 0}
\end{equation}
\begin{equation}
[GM]=[M]^{ 0}[L]^{ 3}[T]^{-2}.
\end{equation}
It is possible to uniquely construct, from these parameters, a quantity $\dot{M}$ 
with dimension $[M]^{ 1}[L]^{ 0}[T]^{-1}$. So, from simple dimensional analysis 
we get the mass accretion rate
\begin{equation}
\dot{M}\sim\overline{\rho}_\infty(GM)^{D-1}(a_\infty)^{3-2D}=C\frac{\rho_\infty}
{l_c^{3(d-1)}}a_\infty\big(\frac{GM}{a^2_\infty}\big)^{D-1}.
\label{massdim}
\end{equation}
Dimension analysis can not be used to fix the dimensionless constant $C$ in the 
above equation and does not give a detailed physical picture. But we have used 
the fractional integrals to derived, in a more detailed analysis in the following 
sections, the mass accretion rate and found that to be consistent with the 
accretion rate derived from the dimensional analysis. 

%%%%%%%%%%%%%%%%%%%%%%%%%%%%%%%%%%%%%%%%%%%%%%%%%%%%%%%%%%%%%%%%%%%%%%%%%%%%%%%
%%%%%%%%%%%%%%%%%%%%%%%%%%%%%%%%%%%%%%%%%%%%%%%%%%%%%%%%%%%%%%%%%%%%%%%%%%%%%%%
\section{Fractional Integrals and Hydrodynamic Equations}
%%%%%%%%%%%%%%%%%%%%%%%%%%%%%%%%%%%%%%%%%%%%%%%%%%%%%%%%%%%%%%%%%%%%%%%%%%%%%%%
%%%%%%%%%%%%%%%%%%%%%%%%%%%%%%%%%%%%%%%%%%%%%%%%%%%%%%%%%%%%%%%%%%%%%%%%%%%%%%%
\label{sec3}
The integrals on network of fractals can be approximated by fractional integrals 
\citep{ren} and the interpretation of the fractional integration is connected 
with fractional mass dimension. We consider the fractional integrals as integrals 
over fractional dimension space (up to a numerical factor) and the fractional 
infinitesimal length for a medium with isotropic mass dimension will be given by
\begin{equation}
d\overline{r}=\frac{1}{l_c^{d-1}}r^{d-1}dr,~~~~~d=D/3<1
\end{equation}
where the constant is chosen to derive the standard expression in the limit 
$d\rightarrow 1$. Note that the infinitesimal area and volume elements in 
this ``fractional continuous'' medium of mass dimension $D=3d$ will be
\begin{eqnarray}
d\overline{A_r}     &=& \frac{1}{l_c^{2(d-1)}}\frac{r^{2(d-1)}}{d^2} r^2 \sin
{\theta} d\theta d\phi \\
d\overline{A_\theta}&=& \frac{1}{l_c^{2(d-1)}}\frac{r^{2(d-1)}}{d} r dr \sin
{\theta} d\phi \\
d\overline{A_\phi}  &=& \frac{1}{l_c^{2(d-1)}}\frac{r^{2(d-1)}}{d} r dr 
d\theta \\
d\overline{V}       &=& \frac{1}{l_c^{3(d-1)}}\frac{r^{3(d-1)}}{d^2} r^2 dr 
\sin{\theta} d\theta d\phi
\end{eqnarray}
and hence the mass enclosed in a sphere of radius $R$ for constant density 
$\rho$ will be
\begin{equation}
M_D=\int_V\rho d\overline{V}=\frac{4\pi\rho}{l_c^{3(d-1)}}\frac{R^{3d}}{3d^3} 
\sim R^D
\label{md}
\end{equation}
which will give the standard expression in the limit $d \rightarrow 1$.

Below we consider the mass density and momentum density balance in such a 
fractal medium of mass dimension $3d$ in the case of accretion onto a central 
gravitating mass $M$ for the case of our interest (i.e. the steady state 
hydrodynamic spherical accretion).

We consider the infinitesimal volume $dV$ in $E^3$ and the balance of mass 
density or the conservation of mass in that volume in the case of steady state 
spherical accretion will imply
\begin{equation}
\frac{d}{dr}[\rho u (r/l_c)^{d-1} d\overline{A_r}] = 0 ~~~\Longrightarrow~~~ \frac{d}{dr}(
\rho u r^{3d-1}) = 0.
\label{continuity}
\end{equation}
This is the generalized mass density balance equation or the generalized 
continuity equation that can be integrated over a surface of constant $r$ to 
give a constant radial accretion rate
\begin{equation}
\dot{M}=\frac{4\pi}{d^2}\rho u l_c^{3(1-d)}{r}^{3d-1}={\rm~Constant~(independent
~of~r)}.
\label{mdot}
\end{equation}

In the infinitesimal volume $dV$ in $E^3$, the balance of momentum density will 
imply 
\begin{equation}
\frac{d}{dt}(\rho {\bf V} d\overline{V}) = {\bf F}^M + {\bf F}^S
\label{vmom}
\end{equation}
where ${\bf V}$ is the velocity vector, ${\bf F}^M$ is the total gravitational 
force acting on the mass contained in the infinitesimal volume $dV$ and ${\bf 
F}^S$ is the total surface force due to pressure acting on the surfaces 
bounding the volume $dV$ and are given by
\begin{equation}
F_r^M = - \frac{GM\rho}{r^2} d\overline{V}
\label{vgr}
\end{equation}
\begin{eqnarray}
F_r^S &=& \widehat{\bf e}_r.(r/l_c)^{d-1}[(pd\overline{A_r}\widehat{\bf e}_r)_r 
          - (pd\overline{A_r}\widehat{\bf e}_r)_{r+dr}+\nonumber \\ 
      & & (pd\overline{A_\theta}\widehat{\bf e}_\theta)_\theta-(pd\overline{A_
          \theta}\widehat{\bf e}_\theta)_{\theta+d\theta} + (pd\overline{A_\phi}
          \widehat{\bf e}_\phi)_\phi-\nonumber \\
      & & (pd\overline{A_\phi}\widehat{\bf e}_\phi)_{\phi+d\phi}]
\end{eqnarray}
and, for spherical accretion where $p=p(r)$, it will be reduced to 
\begin{equation}
F_r^S = - \frac{dp}{dr}d\overline{A_r} dr (r/l_c)^{d-1}.
\label{vpr}
\end{equation}
The total change of radial momentum is given by 
\begin{equation}
\frac{d}{dt}(\rho u d\overline{V})=\frac{\partial}{\partial t}(\rho u 
d\overline{V})+\frac{\partial}{\partial r}[\rho u^2 (r/l_c)^{d-1} d\overline{A_r}]dr.
\label{smom}
\end{equation}
Combining this and equation (\ref{continuity}), in case of steady state we get 
\begin{equation}
\rho u\frac{du}{dr}(r/l_c)^{d-1} d\overline{A_r}dr=- \frac{dp}{dr}d\overline{A_r} 
dr (r/l_c)^{d-1}- \frac{GM\rho}{r^2} d\overline{V}
\label{interm}
\end{equation}
and this can be simplified to the generalized momentum density balance equation
\begin{equation}
u\frac{du}{dr}+\frac{1}{\rho}\frac{dp}{dr}+\frac{GM}{r^2}=0.
\label{momentum}
\end{equation}
This equation, using the equation of state $p=K\rho^\gamma$ ($1 \le 
\gamma \le \frac{5}{3}$) and the boundary condition at infinity, can also be 
integrated to give
\begin{equation}
\frac{1}{2}u^2+\frac{1}{\gamma-1}a^2-\frac{GM}{r}=\frac{1}{\gamma-1}a^2_{\infty}
\label{energy}
\end{equation}
where $a$ is the sound speed given by $a\equiv(dp/d\rho)^{1/2}$ and $a_
{\infty}$ is the sound speed at infinity. This is a local conservation law and, 
as expected, has exactly the same form as that of the continuous medium. 

%%%%%%%%%%%%%%%%%%%%%%%%%%%%%%%%%%%%%%%%%%%%%%%%%%%%%%%%%%%%%%%%%%%%%%%%%%%%%%%
%%%%%%%%%%%%%%%%%%%%%%%%%%%%%%%%%%%%%%%%%%%%%%%%%%%%%%%%%%%%%%%%%%%%%%%%%%%%%%%
\section{Mass accretion rate}
%%%%%%%%%%%%%%%%%%%%%%%%%%%%%%%%%%%%%%%%%%%%%%%%%%%%%%%%%%%%%%%%%%%%%%%%%%%%%%%
%%%%%%%%%%%%%%%%%%%%%%%%%%%%%%%%%%%%%%%%%%%%%%%%%%%%%%%%%%%%%%%%%%%%%%%%%%%%%%%
\label{sec4}

\begin{figure}
\begin{center}
\resizebox*{8.0cm}{8.0cm}{
\includegraphics[angle=-90,width=8.5cm]{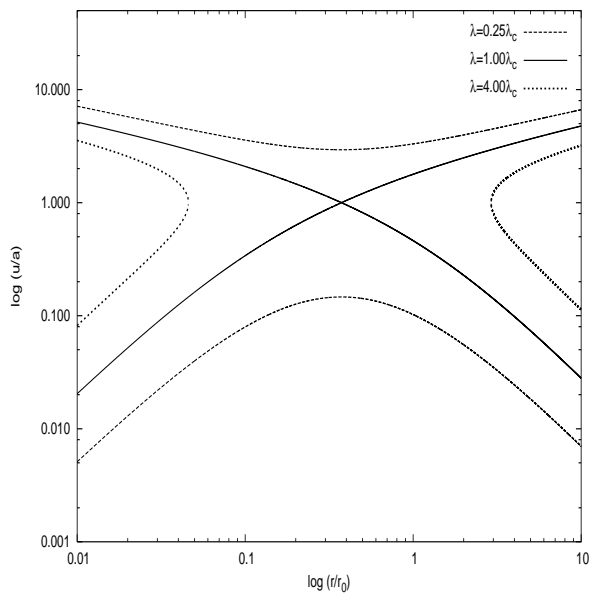}}
\end{center}
\caption{Velocity profile ($u/a$--$r/r_0$, $r_0=GM/a^2_\infty$) for $D=2.55$ and 
$\gamma=\frac{7}{5}$ at $\lambda=0.25\lambda_c$, $1.00\lambda_c$, $4.00\lambda_c$}
\label{figure2}
\end{figure}

We have solved equations (\ref{continuity}) and (\ref{momentum}) for a smooth, 
monotonic solution without any singularities in the flow. Following the standard 
derivation of the ``Bondi solution'', we found that there exist unique eigen 
value solution of the problem for a given $d$ and $\gamma$ and that the solution 
must pass through a critical point where 
\begin{equation}
u_c^2=a_c^2=S(d,\gamma)a^2_\infty,~~S(d,\gamma)=\frac{2}{(6d-1)-3\gamma
(2d-1)}.
\end{equation}
The mass accretion rate may not be steady in a small timescale because of the 
``clumpy'' structure of the medium but a time average mass accretion rate can also 
be calculated following the standard derivation and is given by
\begin{eqnarray}
\dot{M} &=& 4\pi\lambda_c(d,\gamma)\frac{\rho_\infty}{l_c^{3(d-1)}}a_\infty\big
            (\frac{GM}{a^2_\infty}\big)^{3d-1}\nonumber \\
        &=& 4\pi\lambda_c(d,\gamma)\rho_\infty a_\infty(GM/a^2_\infty)^2 F^{-3(1-d)}
\end{eqnarray}
where $F(M,a_\infty,l_c)$ and $\lambda_c(d,\gamma)$ are dimensionless parameters 
\begin{equation}
F(M,a_\infty,l_c)={GM}/{a^2_\infty l_c}
\end{equation}
\begin{equation}
\lambda_c(d,\gamma)=\frac{(3d-1)^{(1-3d)}}{d^2}[S(d,\gamma)]^{\frac{\gamma+1}
{2(\gamma-1)}-(3d-1)}.
\label{lambda}
\end{equation}
This is consistent with the accretion rate derived from the dimensional analysis  
(see Eqn.(\ref{massdim}) in Sec.\ref{sec2}). The solution also uniquely corresponds 
to the physically likely situations properly matching the boundary conditions 
(small velocity at large distance and high velocity at small $r$ in accretion 
flow and the opposite in ``wind flow''). The velocity profile is shown in figure 
(\ref{figure2}) for a particular case where $D=3d=2.55$ and $\gamma=\frac{7}{5}$. 
The general nature of the solution is similar to the ``Bondi solution'' 
\citep{bondi}. Note that the smooth, physically selected solution exists only for 
the transonic case with $\lambda=\lambda_c$. It also corresponds to the maximum 
accretion rate as there exists no meaningful solution either for $\lambda>\lambda_c$ 
or for $\lambda<\lambda_c$.

Note that value of $\lambda_c$ is of order unity. In particular, for $d=1$ and 
$\gamma=\frac{5}{3}$, $\lambda_c=\frac{1}{4}$ \citep{bondi} and for $\gamma=1$, 
$\lambda_c$ is given by
\begin{equation}
\lambda_c(d,\gamma=1)=\frac{(3d-1)^{(1-3d)}}{d^2}e^{3d-\frac{3}{2}}
\end{equation}
which is the limit of equation (\ref{lambda}) as $\gamma \rightarrow 1$, so that 
$\lambda_c$ is continuous at $\gamma=1$.

The most interesting consequence is the dependence of accretion rate on the 
factor $F(M,a_\infty,l_c)$. In the interstellar medium typical ambient sound 
speed (which goes as $\sqrt{T}$) is $\sim10$ km~s$^{-1}$ for a temperature of $\sim10^4$ 
K. In this case the value of $F$ is given by
\begin{equation}
F(M,a_\infty,l_c)=8.907\,\big(\frac{M}{M_\odot}\big)\big(\frac{10~{\rm km~s}}
{a_\infty}\big)^2\big(\frac{1~{\rm AU}}{l_c}\big).
\label{eqnf}
\end{equation}
This is simply the ratio of the length-scale of ``sphere of influence'' of the 
accreting object to the scale length of the fractal structure. As we are mainly 
interested in the medium inside the sphere of influence, our approach to the 
problem of accretion in fractal medium will be valid only if $F\gtrsim1$.

In such situations, this factor can change the mass accretion rate substantially 
and it will crucially depend on the fractional dimensionality $d$. More 
interestingly, for the cases where this factor is important, the accretion rate 
$\dot{M}$ will not be proportional to $M^2$ but will be significantly different 
from that. For example, for $2.5 \lesssim D \lesssim 2.7$ \citep*{nestor} or 
equivalently $0.83 \lesssim d \lesssim 0.90$, the accretion rate $\dot{M} \sim 
M^{\alpha}$ where $1.5 \lesssim \alpha \lesssim 1.7$.

In the actual astrophysical situation, however, the Bondi-Hoyle accretion rate 
will probably be more relevant \citep*[][and references therein]{bhoyle,bondi,
edgar}. \citet{bondi} proposed the interpolation formula \citep*[corrected upto 
a numerical factor by][]{shima} replacing $a_\infty$ by $(a^2_\infty+v^2_\infty)^
{1/2}$ and Bondi radius $r_{\rm B}=GM/a^2_\infty$ by Bondi-Hoyle radius $r_{\rm BH}=
GM/(a^2_\infty+v^2_\infty)$ where $v_\infty$ is the gas velocity relative to the 
star at a large distance. For a medium with density and velocity structure, it 
is not possible to define a unique $v_\infty$ and the present analysis can not be 
extended to such a situation. But it may be extended to a more idealized case where 
the medium has a fractal structure and the accretor is moving in a velocity $v_\infty$ 
relative to the medium at a large distance. Following a similar interpolation method, 
the accretion rate in this case will be 
\begin{equation}
\dot{M}_{\rm BH}=\frac{4\pi\overline{\lambda}_c G^2 \rho_\infty M^2}{(a^2_\infty
+v^2_\infty)^{3/2}} \overline{F}^{\,-3(1-d)}
\label{bhaccr}
\end{equation}
where $\overline{F} \sim r_{\rm BH}/l_c$ and $\overline{\lambda}_c$ is a factor of 
the order of unity. Here we have assumed that the gas velocity relative to the star 
at a large distance will modify the accretion rate in fractal medium in a similar 
way that of in the continuous medium. Though a more detailed analysis is required to 
get the Bondi-Hoyle accretion rate and, in particular, $\overline{F}$ in turbulent 
medium, it is most likely that it will not change our main result that the accretion 
rate will be proportional to $M^{3d-1}$.

%%%%%%%%%%%%%%%%%%%%%%%%%%%%%%%%%%%%%%%%%%%%%%%%%%%%%%%%%%%%%%%%%%%%%%%%%%%%%%%
%%%%%%%%%%%%%%%%%%%%%%%%%%%%%%%%%%%%%%%%%%%%%%%%%%%%%%%%%%%%%%%%%%%%%%%%%%%%%%%
\section{Astrophysical Implications}
%%%%%%%%%%%%%%%%%%%%%%%%%%%%%%%%%%%%%%%%%%%%%%%%%%%%%%%%%%%%%%%%%%%%%%%%%%%%%%%
%%%%%%%%%%%%%%%%%%%%%%%%%%%%%%%%%%%%%%%%%%%%%%%%%%%%%%%%%%%%%%%%%%%%%%%%%%%%%%%
\label{sec5}

\begin{figure}
\begin{center}
\resizebox*{8.0cm}{8.0cm}{
\includegraphics[angle=-90,width=8.5cm]{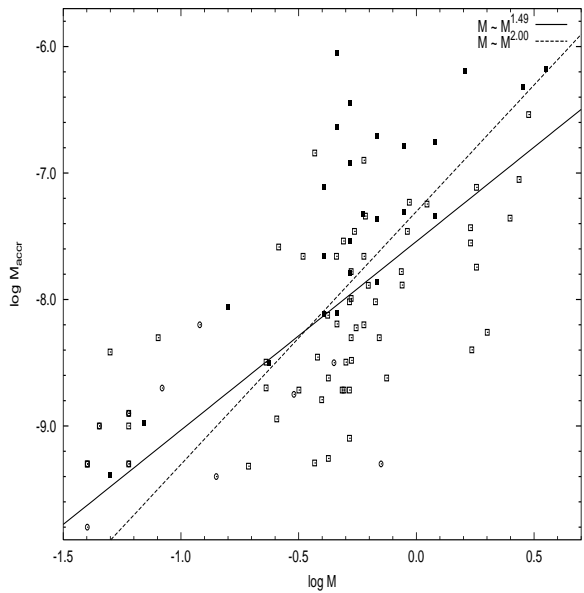}}
\end{center}
\caption{Comparison of data and theory for accretion rates of PMS stars and brown dwarfs. 
The stellar mass is in $M_{\odot}$ and the accretion rate is in $M_{\odot}\,{\rm yr}^
{-1}$}
\label{figure4}
\end{figure}

Our results have important implications for a number of astrophysical problems. One of 
these, for example, is the problem of pre-main-sequence (PMS) accretion. 
Numerical simulation shows that approximating the PMS accretion process as Bondi-Hoyle 
accretion leads to agreement between simulation and observation \citep{padoan}. For a 
solar mass star and for typical sound speed $a_\infty \sim 0.2$ km~s$^{-1}$ in ambient 
molecular cloud filaments \citep{padoan}, a fractal scale length of $\sim 10$ AU will be 
smaller than the size of the sphere of influence and hence our approach should be valid 
in this particular case. Here we compare our central result with observational data and 
numerical simulation results. For that we have taken the accretion rates of PMS stars 
and brown dwarfs compiled by \citet{padoan} \citep*[from][and references therein]{natta,
white,muz} includes all detections but no upper limit. Though the the data do not provide 
any strong support for one theory as opposed to another one due to scatter in the accretion 
rate attributed to (i) an age dependence of the accretion rate, (ii) variation of 
$\rho_\infty$, $a_\infty$ and $v_\infty$ and (iii) interaction of accretion flow with jets 
and outflows on smaller scale \citep{padoan}, we find that our theory is consistent with 
these data. For example, the best fit value of $\alpha$ is $1.49\pm0.13$ for 
$log\,(\dot{M}_{\rm accr}/M_{\odot}\,{\rm yr}^{-1}) \geq -10$. Even when we include 
all the compiled data of \citet{padoan}, the best fit value of $\alpha$ is consistent 
with our theory within $3\sigma$ error for $2.5 \lesssim D \lesssim 2.7$. In figure 
(\ref{figure4}) we have shown the data and the best fit for $log\,(\dot{M}_{\rm accr}/
M_{\odot}\,{\rm yr}^{-1}) \geq -10$. Data from \citet{muz} and \citet{white} are 
shown as filled squares and empty squares respectively and the rest of the data are 
shown as empty circles. We also found that for higher accretion rate and higher central 
mass, the exponent is significantly different from that of the Bondi-Hoyle accretion. 
For a smaller central mass, the self-gravity that we have neglected in our analysis 
may change the accretion rate significantly.

It is possible to use a high resolution simulation \citep[e.g.][]{krumholz} to find 
out the fractional dimensionality and the scale length $l_c$ by using equation (\ref{md}) 
and counting the number of particle $n(R)$ on boxes of different scale length ($R$). On the 
other hand, comparing the result with existing numerical simulation results is not very 
straightforward as in most of the cases it is assumed that $\dot{M}_{\rm accr}=AM^2$ and 
the coefficient $A$ is computed for the assumed $M^2$ dependence on accretion rate. But 
with the existing published simulation results, it is still possible to check if the 
variation of this coefficient $A$ within computational uncertainty can consistently 
accommodate our result and we found that the best fit of the observational data and 
our theoretical estimate are within $1.5 \sigma$ scatter around the computed accretion 
rate of \citet{padoan}. On the other hand, to get a normalized mean accretion rate 
\citep[see][for details]{krumholz} of $0.1$ for a Mach number of $5$ and $D=2.55$, we 
get, from Eqn.(\ref{bhaccr}), $r_{\rm B}/l_c \sim 17740$. For a solar mass star and for a 
typical sound speed $a_\infty \sim 0.2$ km~s$^{-1}$, this translates to $l_c \sim 
1.3$ AU. Fractal structure at this scale is not observationally ruled out. But 
one should take this as an order of magnitude estimate of the scale length as factors 
like velocity structure or magnetic field in real complicated astrophysical situations 
may alter the accretion rate significantly.

The other potential implication is the black hole accretion. The model growth of 
galactic center black holes assume that the black holes accrete at Bondi rate 
\citep*[e.g.,][]{spring}. If the gas around the black hole is fractal in nature, then 
one should rather use the modified accretion rate for fractal medium. Here we would 
like to mention the caveat that there is no definitive observational evidence against 
or for the fractal nature of the medium in this case. Even if the medium is fractal, 
the validity of our approach, as mentioned in \S3, crucially depends on the mass of 
the accretor, the scale length of the fractal and the ambient sound speed.

%%%%%%%%%%%%%%%%%%%%%%%%%%%%%%%%%%%%%%%%%%%%%%%%%%%%%%%%%%%%%%%%%%%%%%%%%%%%%%%
%%%%%%%%%%%%%%%%%%%%%%%%%%%%%%%%%%%%%%%%%%%%%%%%%%%%%%%%%%%%%%%%%%%%%%%%%%%%%%%
\section{Discussion}
%%%%%%%%%%%%%%%%%%%%%%%%%%%%%%%%%%%%%%%%%%%%%%%%%%%%%%%%%%%%%%%%%%%%%%%%%%%%%%%
%%%%%%%%%%%%%%%%%%%%%%%%%%%%%%%%%%%%%%%%%%%%%%%%%%%%%%%%%%%%%%%%%%%%%%%%%%%%%%%
\label{sec6}
We have derived the steady state hydrodynamic equations to describe the spherical 
accretion in a fractal medium by replacing the fractal medium by a ``fractional 
continuous'' model. We have derived the steady state accretion rate in terms of 
boundary conditions at infinity in this simplified situation without considering 
the self-gravity of the material for a medium where the dimensionality is 
independent of position and orientation. Magnetic fields, which we have not 
included in these models, will certainly play a major role in determining the 
dynamics. But even without the inclusion of magnetic field we have got the 
following results 

We have found that there exists a unique solution with maximum mass accretion 
rate and the general nature of the solution is similar to the ``Bondi 
solution'' \citep{bondi} even for a fractal medium with $D=3d<3$. The mass 
accretion rate, in cases, may differ substantially depending on the fractional 
dimensionality. In particular, the accretion rate is likely to be significantly 
different from the ``Bondi accretion rate'' and will be proportional to $M^{D-1}$ 
in case of accretion in fractional medium with scale length $l_c$ very different 
from $GM/a^2_\infty$.

One limitation of our analysis is that we have not considered the self-gravity of 
the medium. This is justified provided the central mass $M$ is very large. In cases 
where self-gravitation is not negligible, it can change the accretion rate significantly. 
A more important limitation is that the present analysis does not account for turbulent 
velocity structure. It will certainly play an important role to determine the accretion 
rate but is probably unlikely to change the mass dependence. In a real astrophysical 
situation with both density and velocity structure in the medium, the analytical mass 
accretion rate will hence not be applicable. Neglecting the magnetic fields, as mentioned 
earlier, is the other major limitation of the analysis. The presence of magnetic field 
is more likely to suppress the accretion rate and magnetohydrodynamic simulation in 
fractal medium will be required to make a more definitive statement. But the central 
result, that the accretion rate will be proportional to $M^{D-1}$ will not be affected 
by the addition of magnetic fields.

Our assumption of the equation of state of the form $p=K\rho^\gamma$ may seems to 
be a considerable limitation but that is not likely to be the case. In ordinary 
Bondi-Hoyle accretion in continuous medium, changing the equation of state changes 
changes the accretion rate by a numerical factor of the order unity \citep*{ruff, 
ruffarn} and hence will not change the result much. A more detail justification 
of the assumption can be found in \citet*{krumholz}.

%%%%%%%%%%%%%%%%%%%%%%%%%%%%%%%%%%%%%%%%%%%%%%%%%%%%%%%%%%%%%%%%%%%%%%%%%%%%%%%
%%%%%%%%%%%%%%%%%%%%%%%%%%%%%%%%%%%%%%%%%%%%%%%%%%%%%%%%%%%%%%%%%%%%%%%%%%%%%%%
\section{Conclusion}
%%%%%%%%%%%%%%%%%%%%%%%%%%%%%%%%%%%%%%%%%%%%%%%%%%%%%%%%%%%%%%%%%%%%%%%%%%%%%%%
%%%%%%%%%%%%%%%%%%%%%%%%%%%%%%%%%%%%%%%%%%%%%%%%%%%%%%%%%%%%%%%%%%%%%%%%%%%%%%%
\label{sec7}

We have shown that the accretion rate onto a mass embedded in a fractal medium 
may differ, in some cases, significantly from the Bondi accretion rate even in the 
simplest situation. We have used the simple model of accretion onto a mass from a 
fractal medium of mass dimensionality $D\leq 3$ and derived a self-consistent 
solution that matches the ``Bondi solution'' as $D \rightarrow 3$. The primary 
result of our investigation is that theoretically accretion rate will be 
proportional to $M^{D-1}$. The observational accretion rate data and the numerical 
simulation for particular astrophysical problem of accretion onto PMS stars is 
consistent with our result. Our findings suggest that the fractal structure of 
the medium around the accreting mass is playing a major role to determine the 
accretion rate and its dependence on the central mass. The agreement of the 
theoretical prediction with existing numerical simulation implies the consistency 
of the approach. A number of previously published numerical and analytical 
results have not considered this effect and may need to be reconsidered. We 
leave a more detailed treatment of the problem, including the effects discussed 
in this work, the effect of self-gravity and stability of fractal accretion, to 
a future work.

%%%%%%%%%%%%%%%%%%%%%%%%%%%%%%%%%%%%%%%%%%%%%%%%%%%%%%%%%%%%%%%%%%%%%%%%%%%%%%%
%%%%%%%%%%%%%%%%%%%%%%%%%%%%%%%%%%%%%%%%%%%%%%%%%%%%%%%%%%%%%%%%%%%%%%%%%%%%%%%
\section{acknowledgments}
%%%%%%%%%%%%%%%%%%%%%%%%%%%%%%%%%%%%%%%%%%%%%%%%%%%%%%%%%%%%%%%%%%%%%%%%%%%%%%%
%%%%%%%%%%%%%%%%%%%%%%%%%%%%%%%%%%%%%%%%%%%%%%%%%%%%%%%%%%%%%%%%%%%%%%%%%%%%%%%
\label{sec8}
We thank Paolo Padoan and Vasily E. Tarasov for useful discussions and Paolo 
Padoan for kindly providing us the compiled data of the accretion rate of PMS 
stars and brown dwarfs. We are grateful to Rajaram Nityananda for reading an 
earlier version and suggesting the dimensional argument as a cross check of the 
calculations. Thanks to Kandaswamy Subramanian for his comments on an earlier 
version of the paper. We also thank Jayaram N. Chengalur, Raghunathan Srianand 
and Ranjeev Misra for useful discussion. We are grateful to the anonymous 
referee for useful comments and for prompting us into substantially improving 
this paper. This research was supported by the National Centre for Radio 
Astrophysics (NCRA) of the Tata Institute of Fundamental Research (TIFR).

~\\

~\\

%%%%%%%%%%%%%%%%%%%%%%%%%%%%%%%%%%%%%%%%%%%%%%%%%%%%%%%%%%%%%%%%%%%%%%%%%%%%%%%
%%%%%%%%%%%%%%%%%%%%%%%%%%%%%%%%%%%%%%%%%%%%%%%%%%%%%%%%%%%%%%%%%%%%%%%%%%%%%%%

\bsp

\label{lastpage}

\end{document}